\documentclass[newstyle,twocolumn,proceedings]{rmaa}
\usepackage{paralist}
\usepackage{psfrag,color}

\newcommand{\der}[2]  { \frac{{\rm d}#1}{{\rm d}#2} }

\renewcommand{\P}[1]{%
\ifnum#1=1\hbox{OW~168--326E}\fi
\ifnum#1=2\hbox{OW~167--317}\fi
\ifnum#1=3\hbox{OW~163--317}\fi
\ifnum#1=5\hbox{OW~158--323}\fi
\ifnum#1=0\hbox{OW~171--334}\fi}
\title{Superbubbles vs  Super-galactic winds}
\author{Guillermo Tenorio-Tagle, \altaffilmark{1},
        Sergey Silich            \altaffilmark{1} 
        \& Casiana Mu\~noz-Tu\~n\'on \altaffilmark{2}}
\altaffiltext{1}{Instituto Nacional de Astrof\'\i{}sica Optica y 
                 Electr\'onica (INAOE), AP 51, 72000 Puebla, M\'exico} 
\altaffiltext{2}{Instituto  de Astrof\'\i{}sica de Canarias (IAC), 
                 E 38200 La Laguna, Tenerife Spain}
\shortauthor{Guillermo Tenorio-Tagle, Silich \& Mu\~noz-Tu\~non}
\shorttitle{Superbubbles vs super-galactic winds}
\keywords{hydrodynamics --- ISM: metal abundances--- ISM:
          mixing  --- Stars: starbursts energetics}

\abstract{%

Here we stress some of the major differences between supergalactic winds
and giant superbubbles evolving into the giant low density haloes of 
galaxies. Both events are the result of massive bursts of star
formation within the densest regions of the host galaxies. However, 
supergalactic winds are able to channel the metals produced by the
recent burst straight into the intergalactic medium while superbubbles 
fail to reach the outskirts of the host galaxies and thus retain the
newly processed metals and with them eventually raise the abundance of 
their ISM. We review the properties of major bursts of star formation
and the critical energy (and mass of the starburst) required for mass 
ejection both in the case of an ISM strongly flattened by rotation
into a thin disk and that imposed by a more extended ISM distribution 
arising from a smaller rotation. The limits are thus establish for
galaxies with an ISM mass in the range 10$^6$ M$_\odot$ to more than 
10$^9$ M$_\odot$, and are compared with a sample of local galaxies.
Some of these galaxies seem to be above the critical limit despite the 
fact that their structure is clearly that of a superbubble. True 
supergalactic winds, as evidence by M82, are shown to exceed the
critical limit by more than an order of magnitude and thus the 
limit derived by Silich \& Tenorio-Tagle (2001) for mass ejection 
should be regarded as a lower limit.}
\resumen{
Se revisan las diferencias f\'\i{}sicas que existen entre
superburbujas y super vientos gal\'acticos. Los dos eventos resultan
de la energ\'etica producida por grandes brotes de formac\'on estelar.
Sin embargo los super vientos gal\'acticos permiten el escape de los 
metales recien procesados por el brote estelar hacia el medio
intergal\'actico, mientras que las superburbujas al no llegar a
alcanzar el borde de su galaxia anfitriona retienen a los nuevos
metales con los que m\'as tarde enriquecer\'an  su medio
interestelar. Se evaluan por lo mismo las principales propiedades de 
brotes masivos de formaci\'on estelar as\'\i{} como   
la tasa m\'\i{}nima de energ\'\i{}a mec\'anica que un starburst ha de 
producir para causar la expulsi\'on de su material reci\'en 
procesado fuera de su galaxia anfitriona, tanto en el caso de galaxias 
con alta rotaci\'on que presenten un disco aplanado, como
para el caso de galaxias con una rotaci\'on menor que imponen un 
l\'\i{}mite ordenes de magnitud m\'as energ\'etico. 
Se establecen los l\'\i{}mites para galaxias con una masa de material 
interestelar que va de 10$^6$ M$_\odot$ a m\'as de 10$^9$ M$_\odot$ y 
estos son comparados con un grupo de galaxias locales. Algunas de
estas galaxias aparecen por encima del l\'\i{}mite cr\'\i{}tico de 
ejecci\'on de masa a pesar de que su estructura es la predicha 
te\'oricamente como  t\'\i{}pica de una superburbuja. Se muestra 
tambi\'en que verdaderos supervientos gal\'acticos, como es el caso 
de M82, exceden el l\'\i{}mite cr\'\i{}tico por m\'as de un orden de 
magnitud, lo que implica que el \'\i{}mite establecido por 
Silich \& Tenorio-Tagle (2001) es una cota inferior.}
\listofauthors{W.~J.~Henney \& S.~J.~Arthur}
\indexauthor{Henney, W.~J.}
\indexauthor{Arthur, S.~J.}
\begin{document}

\maketitle
%%
%% And here starts the text....
%%
\section{Introduction}
\label{sec:intro}

The development of super-galactic winds (SGWs) is intimately related
to the properties of the central massive starburst and to the 
distribution of interstellar matter (ISM) in the host galaxy. 
Clearly, to develop a SGW, through which the newly processed metals 
from the starburst are directly injected into the intergalactic medium 
(IGM), a channel, a free path,  has to be carved into the ISM to
finally connect the starburst with 
the medium surrounding the host galaxy.  The process involves the 
propagation of shock waves into the disk and the halo of the galaxy,
which lead to the formation and evolution of the so called superbubbles. 
These have often been mistaken with SGWs, despite their 
unique appearance which at all wavelengths is completely different to 
that presented by SGWs. Here we stress some of the main differences 
between SGWs and superbubbles and center our attention on three
important issues: 1) The expected properties of massive burst of star 
formation. 2) The development of superbubbles, and 3) the physical 
properties of SGWs. Especial emphasis is made on the physics 
applicable to SGWs driven by the most massive and powerful starbursts 
thought to exist in the Universe. 

\section{The properties of massive bursts of stellar formation}

Our knowledge of stellar evolution has now been assembled by several 
groups in order to predict the properties of stellar clusters, given 
an IMF and a stellar mass range. These are the so called synthesis models 
of starbursts (Mas-Hesse and Kunth 1991, Leitherer \& Heckman 1995) 
which predict a variety of observable quantities, as well as the 
energetics that one is to expect from a stellar cluster, as a function 
of time. Now we know that a  10$^6$ M$_{\odot}$ coeval starbursts with 
a Salpeter IMF and  stellar masses in the range 1 - 100 M$_\odot$ 
leads to  the appearance of several thousands of O stars strongly 
correlated in space (within a radius much smaller than 100 pc).
All massive stars undergo strong stellar winds and all of them with a 
mass larger than 8 M$_\odot$ will end their evolution exploding as 
supernova. And therefore, one is to expect from our hypothetical
cluster several tens of thousands of SN over a time span of some 40 Myr. 
During the supernova phase a 10$^6$ M$_\odot$ stellar cluster will 
produce an almost constant energy input rate of the order of 
10$^{40}$ erg s$^{-1}$. On the other hand, the ionizing luminosity 
emanating from the 10$^6$ M$_\odot$ cluster would reach a constant 
value of 10$^{53}$ $UV$ photons s$^{-1}$ during the first 3.5 Myr of 
evolution to then drastically drop (as t$^{-5}$) as the most massive 
members of the association explode as supernova. The rapid drop in the 
ionizing photon flux implies that after 10 Myr of evolution, the $UV$ 
photon output would have fallen by more than two orders of magnitude 
from its initial value and the HII region that they may have 
originally produced would have drastically reduced its dimensions.
Thus the HII region lifetime is restricted to the first 10 Myr of 
the evolution and is much shorter than the supernova phase. It is 
important to realize that only 10$\%$ of the stellar mass goes into 
stars with a mass larger than 10 M$_\odot$, however, it is this   
10$\%$ the one that causes all the energetics from the starburst.
Being massive, although smaller in numbers, they also reinsert into the 
ISM, through their winds and SN explosions, almost 40$\%$ of the 
starburst original mass. And thus from a starburst with 
an initial mass of 10$^6$ M$_\odot$ one has to expect a total of 
almost 4 $\times 10^5$ M$_\odot$ violently injected back into the ISM,
during the 4 $\times 10^7$ years that the SN phase lasts. From these, 
almost 40,000 M$_\odot$ will be in oxygen ions and less than 
1000 M$_\odot$ in iron (see Silich et al. 2001).
 
One of the features of the stellar synthesis models regarding the 
energetics of star clusters is that they fortunately scale linearly
with the mass of the starburst. It is therefore simple to derive the 
properties of starbursts of different masses, for as long 
as they present the IMF, metallicity and stars in the same mass range 
considered by the models.

There is a growing observational evidence pointing at massive,
centrally condensed congregations of stars,
as the fundamental unit of massive star formation (see Ho, 1997). 
These luminous structures often referred to as young globular clusters 
or super-star clusters, present a
mass in the range of  10$^5$ to a few times 10$^6$ M$_\odot$ in stars, 
all pulled together within a typical radius $R_{SB}$ 
$\sim$ 3 pc. The brightest ones have luminosities up to two orders of 
magnitude higher than R136 in 30 Doradus. Similar super-star cluster
properties have been inferred from HST-STIS observations of AGN 
(see Colina et al. 2002), and 
from radio continuum measurements of ultracompact HII regions not 
visible in optical images, fact that points to the youngest, densest 
and most highly obscured star formation events ever found 
(see Kobulnicky \& Johnson 1999; Johnson et al. 2001).   
The massive concentrations imply a high efficiency of star formation 
which permits even after long evolutionary times  the  
tight configuration that characterizes them, despite stellar evolution and its 
impact through photo-ionization, winds and supernova explosions,
believed to efficiently disperse the gas left over from star formation.
It is thus the self-gravity that results from the high efficiency what 
keeps the sources bound together. 

The close spacing between the super-star cluster sources warrants a 
very efficient thermalization of all their winds and supernova 
explosions, leading to the high central overpressure that is to drive 
both the superbubble and in some cases the SGW. The outflow is fully 
defined by three quantities: the mass and mechanical energy deposition 
rates (hereafter $\dot M_{SB}$ and $\dot E_{SB}$) and the radius that 
encompasses the newly born sources ($R_{SB}$).

The total mass and energy deposition rates define the central temperature
and thus the sound speed $c_{SB}$ 

%---------------------------------------------------------------------- 
\begin{equation}
      \label{eq.01} 
T_{SB} = \frac{0.299 \mu}{k} \frac{\dot E_{SB}}{{\dot M}_{SB}} , 
\end{equation} 
%----------------------------------------------------------------------

\noindent 
where $\mu$ is the mean mass per particle and $k$ the 
Boltzmann constant. On the other hand, the density of matter streaming 
out of $R_{SB}$ is:

%---------------------------------------------------------------------- 
\begin{equation}
\rho = \frac{{\dot M_{SB}}}{4 \pi R_{SB}^2 c_{SB}} ,
\end{equation} 
%----------------------------------------------------------------------

Thus at $R_{SB}$  (see Chevalier \& Clegg 1985; hereafter CC85), the 
ratio of thermal and kinetic energy flux to the total flux is

%---------------------------------------------------------------
\begin{equation}
      \label{eq.02} 
F_{th}/F_{tot} = \frac{\frac{1}{\gamma - 1}\frac{P}{\rho}}
                 {\frac{u^2}{2} + \frac{\gamma}{\gamma - 1}
                 \frac{P}{\rho}} = \frac{9}{20}
\end{equation} 
\begin{equation}
      \label{eq.03}   
F_{k}/F_{tot} = \frac{u^2/2}{\frac{u^2}{2} + \frac{\gamma}{\gamma - 1}
                 \frac{P}{\rho}} = \frac{1}{4} .
\end{equation} 
%-------------------------------------------------------------
There is however a rapid evolution as matter streams away from the
central starburst. After crossing $r = R_{SB}$ the gas is immediately 
accelerated by the steep pressure gradients and rapidly reaches 
its terminal velocity ($V_t \sim 2 c_{SB}$). This is due to a fast 
conversion of thermal energy, into kinetic energy of the resultant 
wind.

In a recent communication (Silich et al. 2003), we have revised the 
properties of SGWs by solving the flow equations dropping the
assumption of an adiabatic flow made by Chevalier \& Clegg (1985). In 
this case, the steady-state solution results from solving 

%---------------------------------------------------------------
\begin{eqnarray}
      \label{eq.1a}
      & & \hspace{-0.5cm}
\frac{1}{r^2} \der{}{r}\left(\rho u r^2\right) = 0 ,
      \\[0.2cm]
      \label{eq.1b}
      & & \hspace{-0.5cm}
\rho u \der{u}{r} = - \der{P}{r} ,
      \\[0.2cm]
     \label{eq.1c}
      & & \hspace{-0.5cm}
\frac{1}{r^2} \der{}{r}{\left[\rho u r^2 \left(\frac{u^2}{2} +
\frac{\gamma}{\gamma - 1} \frac{P}{\rho}\right)\right]} = - Q,
\end{eqnarray}
%-------------------------------------------------------------
where $Q$ is the cooling rate ($Q = n^2 \Lambda$), $n$ is the wind number 
density and $\Lambda$ is the cooling function. 
The main effect is to largely reduce the size of the X-ray emitting zone,
particularly in the case of powerful and compact starbursts.

The energy dumped by the central starburst, is to cause a major impact
on the surrounding gas. The supersonic stream leads immediately 
to a leading shock able to heat, accelerate and sweep all the
overtaken material into a fast expanding shell. 
In this way, as the free wind takes distance to the star cluster
boundary, its density, temperature and thermal pressure will
drop as $r^{-2}$, $r^{-4/3}$ and  $r^{-10/3}$, respectively (CC85).

Note however that such a flow is exposed to the appearance of reverse 
shocks whenever it meets an obstacle cloud or when its thermal pressure   
become lower than that of the surrounding gas, as it is the case within
superbubbles. There, the high pressure acquired by the swept up ISM 
becomes larger than that of the freely expanding ejecta (the free wind 
region; FWR), where $\rho$, $T$ and $P$ are rapidly falling.
The situation rapidly causes the development of a reverse shock, the 
thermalization of the wind kinetic energy  
and a much reduced size of the FWR. Thus for the FWR to extend 
up to large distances away from the host galaxy, the shocks  would have had
to evolve and displace all the ISM, leading to a free path into the 
intergalactic medium and to a supergalactic wind
with properties (density, temperature and thermal pressure) in
principle similar to those derived by CC85 for a free wind.  

\section{The Properties of superbubbles}

Given the extreme violence of the ejection process, either through 
supernova explosions or strong stellar winds,
the presence of strong reverse shocks assures that upon the 
thermalization the ejected matter would reach temperatures ($T
\sim 10^7 - 10^8$ K) that would strongly inhibit recombination and 
thus the detection of the newly processed material at optical
frequencies. Furthermore, it is now well understood 
that it is this hot high pressure gas the one that fills the interior of
superbubbles and that drives the outer shock that sweeps and
accelerates the surrounding ISM. 
It has then become clear in recent years that the metallicity detected 
in blue compact dwarfs, the same as in all other
galaxies, results from their previous history of star formation and 
has nothing to do with the metals presently ejected by their powerful 
starbursts. The continuous energy input rate that in the coeval
starburst model lasts until the last 8 M$_\odot$ star explodes as 
supernova ($t \sim 4 \times 10^7$ yr) or it extends until the star 
formation phase is over in the continuous star formation model, 
reassures that the high temperature of the ejected matter is
maintained above the recombination limit ($T \sim 10^6$ K) allowing 
superbubbles to reach dimensions in excess of 1 kpc. During this 
phase the first step towards mixing takes place. 
About 10$\%$ of the interstellar matter swept up and 
stored in the expanding supershell, 
becomes thermally evaporated and thus injected into the
superbubble interior during the 
evolution. As the evaporated matter streams into the 
superbubble it acquires a similar temperature to that 
of the supernova ejected gas.
Under such conditions mixing is expected to become a rapid process. 
This is because the large temperatures  favor diffusion,
and also because the large sound speed ensures an
efficient stirring between the two gases. Mixing of the evaporated ISM 
with the ejected metals lowers the metallicity of the superbubble 
interior. Silich et al.(2001) 
have predicted the values to be expected from X-ray observations 
during this phase if one uses either iron or oxygen as tracers.
There is however, no diffusion of the highly metal enriched 
superbubble gas over the matter either in the expanding shell 
or its immediate surroundings (see Tenorio-Tagle 1996, Oey 2003). For 
a true enrichment of the ISM one would have to wait 
for cooling and the large-scale dispersal of the newly processed 
metals. This may take a few times 10$^8$ years.  

The  high internal pressure ($P_{int}$) resultant from the
thermalization of the fast ejecta at the reverse shock drives the 
outer shock to collect the surrounding ISM of density ($\rho$) into a 
dense shell. This in a constant density medium leads to a continuous 
deceleration of the remnant. However, if the evolution takes place in 
a disk-like configuration, and the density falls off steeply in the 
direction perpendicular to the plane
of the galaxy, much more rapidly than the fall of the pressure, then 
the shock will be force to accelerate in the direction of the 
density gradient. This is the moment of breakout (Kompaneets 1960).
Upon breakout, the section of the shell following the shock will also 
experience a sudden acceleration, fact that will promote
the development of Rayleigh - Taylor instabilities that will lead to 
its fragmentation. 
The hot gas filling the remnant interior will then be able to stream 
between fragments, venting the high pressure of the remnant either 
into the extended halo of the galaxy, forming a large superbubble,
or into the IGM in the case of 
a flattened disk-like system, leading eventually to a SGW. In both 
cases, the evolving remnant would appear more and more 
displaced from the galaxy disk, fact that has lead to the confusion 
between superbubbles and SGWs.

Take for example the case of NGC 1569 thoroughly reviewed recently by 
Martin et al. (2002). There the X-ray emission detected by Chandra,
clearly looks displaced from the HI disk. The authors found a 
starburst age of $\sim 10 - 20 \times 10^6$ yr. Also, using oxygen as 
tracer, they ascribed a metallicity to the X-ray cloud of at least 
2 Z$_\odot$. This implies a total oxygen content of some 34,000 
M$_\odot$. All of these results are in excellent agreement with the 
results of the evolution of superbubbles obtained by Silich et al. (2001). 

NGC 1569 is clearly not undergoing a supergalactic wind. This issue was also 
addressed by Martin et al., who noticed that the X-ray emissivity is 
fairly even, as expected in superbubbles, instead of stratified, given 
the rapid drop in density and temperature away from the central source,
as predicted for SGWs. Furthermore, they also noticed that the X-ray 
emission implies an interaction with the halo of the 
galaxy, even in the locations where the HI observations have failed 
to detect it. An important issue that may again lead to confusion is
the fact that NGC 1569 with a total mass of a few 
$\times 10^8$ M$_\odot$ is supposed to have a mechanical energy input 
rate above the threshold required for mass ejection into the IGM 
(see Silich \& Tenorio-Tagle 2001). The same appears to be the case  
for other massive galaxies. This could, in principle, be used as an 
argument in favour of  an eventual development of a SGW. Note however,
that although the threshold imposed by the existence of a galaxy halo 
has brought the threshold up to three orders of magnitude
higher than that envisaged for fattened galaxy disks 
(see Mac Low \& Ferrara 1999), it is still a lower limit to mass ejection.

\section{the energy limit for mass ejection into the IGM}

%-----------------------------------------------------------------------
\begin{figure*}
  \includegraphics[width=\textwidth]{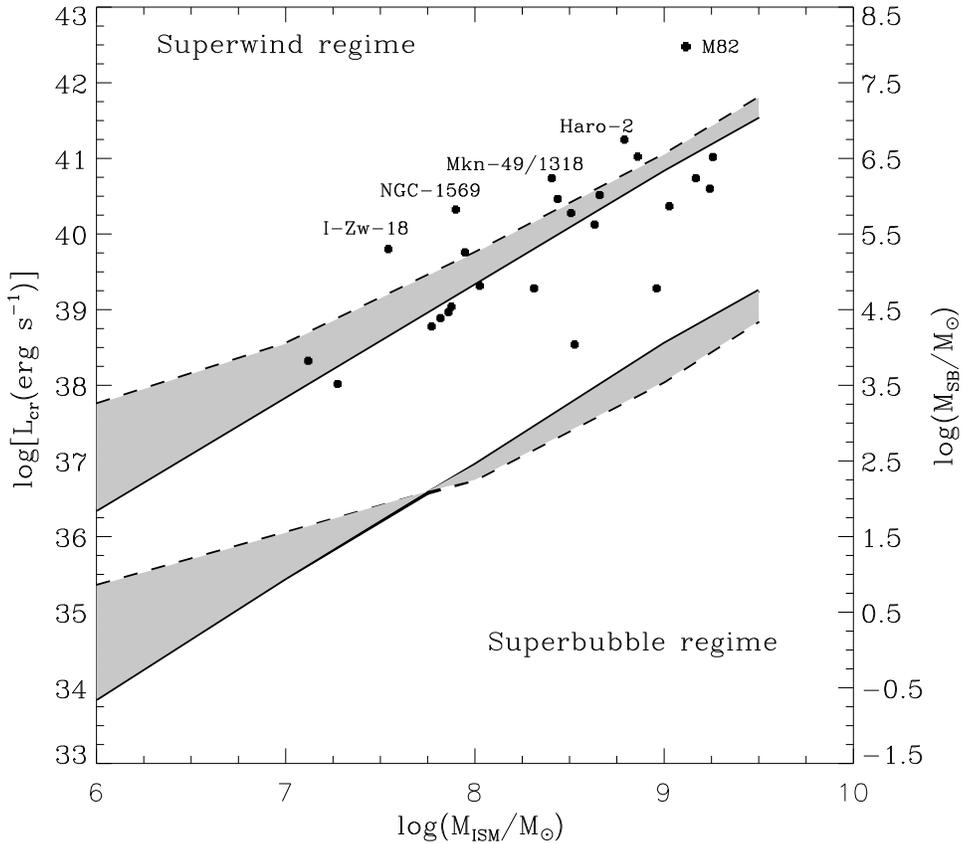}
  \caption{Numerical energy estimates. The log of the critical mechanical 
luminosity (left axis), and of the starburst mass (right-hand side
  axis), required to eject matter from galaxies 
with a $M_{ISM}$ in the range 10$^6$ -- 10$^9$ M$_\odot$. The lower limit 
estimates are shown for galaxies with extreme values of $\epsilon$
(= 0 and 0.9) and for two values of the intergalactic pressure  
$P_{IGM}/k$ = 1 cm$^{-3}$ K (solid lines) and $P_{IGM}/k$ = 100 cm$^{-3}$ K 
(dashed lines). The resolution of our numerical search is 
$ \Delta log L_{cr}$ = 0.1. Each line should be considered 
separately as they divide the parameter space into two distinct 
regions: a region of no mass loss that is found below the line and 
a region in which blowout and mass ejection occur that is found above 
the line. Figure adapted from Silich \& Tenorio-Tagle (2001).
The location of several local dwarf galaxies as studied by 
Legrand et al. 2001 and that of M82 are indicated in the figure.
}
\label{fig:energy}
\end{figure*}
%-----------------------------------------------------------------------

The energy input rates derived by Silich \& Tenorio-Tagle (2001), 
despite the fact of being orders of magnitude above the limits derived 
by Mac Low \& Ferrara (1999), are lower limits to the amounts
required for expelling matter from a galaxy. Particularly because only one 
component of the ISM was considered and because the central densities
adopted are well below the values expected for the star forming cloud 
where the starburst originated. Our estimates thus neglect the effect
of the starburst plowing into the parental cloud  material. 
These are lower limits also because we adopted a constant energy input rate
(see Strickland \& Stevens 2000; Silich et al. 2002) and because our 
approach neglects  the 
presence of a magnetic field which  could also inhibit expansion 
(Tomisaka 1998). It is thus not surprising that some galaxies like
NGC 1569, Haro 2, IZw18 and Markarian 49 (see Legrand et al. 2001), 
lie slightly above the ejection limit
while their physical structure is clearly that of bound superbubbles. 
However, note that this is not the case for M82, where with an energy 
input rate of 3 $\times$ 10$^{42}$ and an ISM mass around 
10$^9$ M$_\odot$, lies more than an order of magnitude above the 
limit prescribed by Silich \& Tenorio-Tagle (2001) and is thus a 
true example of a supergalactic wind. 
 
The indisputable presence of metals (in whatever abundance) in galaxies
implies that the supernova products cannot be lost in all cases. Note in 
particular that many well known disk galaxies have a high 
metal abundance and a large number of centers of star formation. Most
of these exciting star clusters are more massive than the 
$10^4$ M$_\odot$ lower limit established by Mac Low \& Ferrara (1999) 
and Silich \& Tenorio-Tagle (2001)
as the minimum starburst mass required to cause mass ejection 
in the case of  disk-like systems. This lower limit for disk-like galaxies 
with M$_{ISM}\leq$ 10$^9$ M$_\odot$ (see Figure 1) implies that starbursts 
even smaller than the Orion cluster would break through the galaxy outer 
boundary and eject their supernova products into the intergalactic medium.
Nevertheless, galaxies can avoid losing all their freshly 
produced metals by having a halo component, neglected in former studies, 
that acts as the  barrier to the loss of the new metals.

The extended gaseous haloes, despite acting as the barrier to the loss 
of the new metals,
have rather low densities ($<n_{halo}> \leq 10^{-3}$ cm$^{-3}$) 
and thus have a long recombination time ($t_{rec} = 1/(\alpha n_{halo}$); 
where $\alpha$ is the recombination coefficient)
that can easily exceed the life time of the  HII region ($t_{HII}$ = 10$^7$ yr)
produced by the starburst. In such a case, the haloes may remain 
undetected at radio and optical frequencies (see Tenorio-Tagle et al. 
1999, Martin et al. 2002), until large volumes are collected into 
the expanding supershells. Note that the continuous $\Omega$ shape 
that supershells present in a number of galaxies, while remaining 
attached to the central starburst, and their small expansion velocity
(comparable or smaller than the escape velocity of their host galaxy) 
imply that the mechanical energy of the star cluster is plowing into a 
continuous, as yet undetected medium. Supershells crossing the outer 
galaxy boundary into the IGM, should become Rayleigh Taylor 
unstable and rapidly fragment. This will then favour the streaming of 
the hot superbubble interior and thus the development of a
supergalactic wind. Most galaxies in the local universe, including 
those with a low metal abundance, however, do not
present as in M82, a clear evidence of having evolved into such a phase.

\section{Concluding remarks}

- Superbubbles, bound by a large-scale expanding supershell, are
powered by the thermalized ejected matter and thus present a very even 
temperature distribution. On the other hand, supergalactic winds,
given their stratification in density and temperature aught to present a
stratified X-ray emission. 

- The metallicity within the interior of superbubbles changes rapidly 
with time. It exceeds $Z_\odot$, particularly within the first 10 - 20 
Myr of the evolution.

- The development of supergalactic winds depends drastically on the 
distribution of ISM both in the host galaxy disk and in the
halo. Apart from M82, there is no strong evidence of a supergalactic 
wind in the Local Universe.

-Supergalactic winds driven by compact and powerful starbursts are 
subjected to strong radiative cooling, which modifies their
temperature distribution and thus their X-ray appearance.
Radiative cooling reduces drastically the size of the extended X-ray 
emitting zone favouring instead the formation of a rapidly expanding 
photoionized envelope. This may show up as low intensity broad
emission lines associated with luminous starbursts.

\acknowledgements  Financial
support for this research has been provided by  CONACyT, Mexico 
(project number 36132-E), and the Consejo Superior de Investigaciones 
Cient\'\i{}ficas, Spain (grant AYA2001 - 3939).

%% When using the rmaacite package, the \bibitem command should be of
%% the format: 
%%
%%             \bibitem[AUTHOR<YEAR>]{KEY} 
%%
%% so that the \cite{KEY}, etc. commands will work properly. 
%% 
%% If you are doing the citations manually, then you can just use
%% `\bibitem{}' instead. This will give you a warning about
%% `multiply-defined labels' which you can safely ignore.
%% 


\begin{thebibliography}
\bibitem{1} Chevalier, R.A. \& Clegg, A.W. 1985, Nature, 317, 44 (CC85)
\bibitem{2} Colina, L., Gonzalez-Delgado, R., Mas-Hesse, M. \& 
            Leitherer, C. 2002 ApJ 579, 545
\bibitem{3} Ho, L. C. 1997, Rev.MexAA, Conf. Ser. 6, 5
\bibitem{4} Johnson, K. E., Kobulnicky, H. A., Massy, P.  \&  
            Conti, P. S. 2001, ApJ 559, 864
\bibitem{4} Kobulnicky, H. A. \& Johnson, K. E. 1999, ApJ 527, 154   
\bibitem{5} Kompaneets, A.S. 1960, Soviet Phys. Doklady, 5, 46
\bibitem{6}  Legrand, F., Tenorio-Tagle, G., Silich, S., Kunth, D. \& 
             Cervi\~no, M. 2001, ApJ 560, 630 
\bibitem{7}  Leitherer, C. \& Heckman, T.M. 1995, ApJS, 96, 9 
\bibitem{8}  Mac Low, M-M. \& Ferrara, A. 1999, ApJ, 513, 142
\bibitem{9}  Martin, C.L., Kobulnicky, H.A. \& Heckman, T.M., 2002,
             ApJ, 574, 663  
\bibitem{10} Mas-Hesse, J.M. \& Kunth, D., 1991, Astron. Astrophys.
             Suppl. Ser. 88, 399
\bibitem{11} Oey, S., 2002, Astro-ph/0211344 
\bibitem{12} Silich, S. \& Tenorio-Tagle, G. \& Mu\~noz Tu\~non,
             C. 2003, ApJ (in press) 
\bibitem{13} Silich, S. \& Tenorio-Tagle, G. 2001, ApJ 552, 91
\bibitem{14} Silich, S., Tenorio-Tagle, G,  Terlevich, R.  
             Terlevich, E. \& Netzer, H. 2001, Mon. Not. Roy. Ast. 
             Soc., 234, 191
\bibitem{15} Silich, S., Tenorio-Tagle, G,  Mu\~noz Tu\~non, C. \&  
             Cairos, L. M. 2002, AJ, 123, 2438
\bibitem{16} Strickland D.K. \& Stevens I.R. 2000, Mon. Not. Roy. Ast. 
             Soc., 314, 511
\bibitem{17} Tenorio-Tagle, G. 1996, AJ. 111, 1641
\bibitem{18} Tenorio-Tagle, G, Silich, S., Kunth, D., Terlevich, E. \& 
             Terlevich, R. 1999, Mon. Not. Roy. Ast. Soc., 309, 332
\bibitem{19} Tomisaka, K. 1998, Mon. Not. Roy. Ast. Soc., 298, 797  
%%% You can use the following command to manually balance the columns
%%% on the final page. This is best left to the editors, however, and
%%% certainly should be the last thing you do to the paper since any
%%% subsequent editing will likely mess it up. 
%\adjustfinalcols

\end{thebibliography}
\end{document}